\documentclass[prc,twocolumn,secnumroman,tightenlines,amssymb,aps,nobibnotes,
               superscriptaddress,showpacs,balancelastpage,floatfix,nofootinbib]{revtex4}
\usepackage{graphicx}
\usepackage{multirow}           
\usepackage{color}	
\usepackage[utf8]{inputenc}
\expandafter\ifx\csname package@font\endcsname\relax\else
\expandafter\expandafter
\expandafter\usepackage
\expandafter{\csname package@font\endcsname}
\fi
\DeclareRobustCommand\substyle{\name@idx{document substyle}}
\DeclareRobustCommand\classoption{\name@idx{document class option}}
\DeclareRobustCommand\classname{\name@idx{document class}}
\def\name@idx#1#2{{\ttfamily#2}
\index{#2\space#1=\string\ttt{#2}\space#1}\index{#1>#2=\string\ttt{#2}}}

\begin{document}

\title{Influence of fusion dynamics on fission observables: A multi-dimensional analysis}

\author{C. Schmitt}
\email{christelle.schmitt@iphc.cnrs.fr}
\affiliation{Institut Pluridisciplinaire Hubert Curien, 23 rue du Loess, B.P.\,28, 67037 Strasbourg Cedex 2, France}
\author{K. Mazurek}
\affiliation{The Niewodnicza\'nski Institute of Nuclear Physics - PAN, 31-342 Krak\'ow, Poland}
\author{P. N. Nadtochy}
\affiliation{Omsk State Technical University, Mira prospekt 11, Omsk, 644050, Russia}

\date{\today}

\begin{abstract}
\noindent
An attempt to unfold the respective influence of the fusion and fission stages on typical fission observables, and namely the neutron pre-scission multiplicity, is proposed. A four-dimensional dynamical stochastic Langevin model is used to calculate the decay by fission of excited compound nuclei produced in a wide set of heavy-ion collisions. The comparison of the results from such a calculation and experimental data is discussed, guided by predictions of the dynamical deterministic HICOL code for the compound-nucleus formation time. While the dependence of the latter on the entrance-channel properties can straigthforwardly explain some observations, a complex interplay between the various parameters of the reaction can occur in other cases. A multi-dimensional analysis of the respective role of these parameters, including entrance-channel asymmetry, bombarding energy, compound-nucleus fissility, angular momentum and excitation energy, is proposed. It is shown that, depending on the size of the system, apparent unconsistencies may be deduced when projecting onto specific ordering parameters. The work suggests the possibility of delicate compensation effects in governing the measured fission observables, thereby highlighting the necessity of a multi-dimensional discussion.
\end{abstract}
\maketitle

\section{Introduction}

Fission has shown to be a relevant mechanism to learn about a wide spectrum of fundamental nuclear properties, as it is driving by  both reaction mechanism and nuclear structure aspects \cite{krappe:2012, schmidt:2016}. The importance of fission for applications is obvious also, including nuclear power production, transmutation of waste, medicine, network calculations in astrophysics.At intermediate and high excitation energy (above about 40 MeV) the process is a particularly appropriate laboratory for probing nuclear dynamics.\\
Fission is a very entangled process though. Accordingly, a robust understanding has not been reached yet, and various interpretations of the same experimental data exist (see Refs.~\cite{lestone:2009, schmitt:2014, mahata:2017} and therein). Even under "clean" conditions, satisfying Bohr's hypothesis about the independence of the decay of the excited system on the way it was produced, inconsistencies are found (see the discussion in Ref.~\cite{schmitt:2010}). Un-ambiguous insight into the process is complex to extract from heavy-ion-induced fission. Indeed, in this approach, the first stage of the reaction - during which the two ions fuse and produce the excited compound nucleus (CN), can notably affect the subsequent decay: In addition to the obvious role of the fusion stage in determining the excitation energy ($E^*$) and angular momentum ($L$) of the CN, the dynamics of the fusion mechanism itself can play a non-negligible role. The evolution of the composite system formed by the projectile and the target on the way to a fully equilibrated compact system depends on the entrance-channel asymmetry $\alpha$=($A_t$-$A_p$)/($A_t$+$A_p$) (see Ref.~\cite{kumar:1994} and therein). Above the Businaro-Gallone point $\alpha_{BG}$ \cite{businaro:1955} fusion takes place in terms of absorption of the projectile by the target (more generally, the lightest by the heaviest reactant). In other words, the heavy partner takes up the nucleons from the light one, and a mono-nuclear CN is formed in a short time. On the other hand, for more symmetric reactions where $\alpha <  \alpha_{BG}$, a neck develops between the reactants. Projectile and target first form a di-nuclear system which equilibrates in various degrees of freedom; nucleons are exchanged through the neck, and formation of a compact CN takes longer.

\begin{figure*}[!hbt]
\hspace{-0.1cm}
\includegraphics[width=18.cm, height=6cm,angle=0]{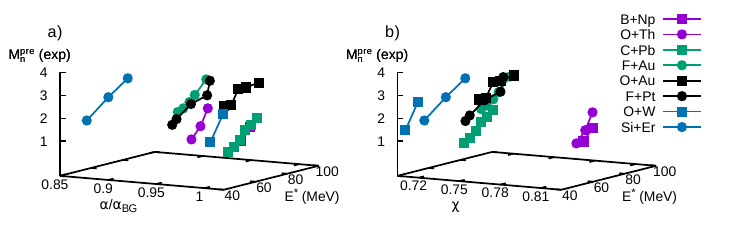}
\vspace{-.60cm}
\caption{(Color online) Correlation between experimental neutron pre-scission multiplicities $M_n^{pre}$ with a) $\alpha$/$\alpha_{BG}$ and $E^*$, and with b) $\chi$ and $E^*$. For experimental points, we refer to Table 1. Error bars are omitted for clarity. Lines connect different energy points for a given reaction. The same color is employed for reaction couples leading to the same CN.} 
\label{fig1}
\end{figure*}

The possible influence of the fusion stage has to be kept in mind whenever experimental data are used to extract fission time scales. In a seminal work in the 90's, Saxena et al. \cite{saxena:1993} analyzed a wide set of projectile-target combinations, with the goal to unfold the different contributions to the total fusion-fission time. In particular, it was attempted to separate the contribution of the fusion stage (CN formation time, $t_{fo}$) from the time scale of the sole fission process. The discrimination is crucial also for further proper unfolding of the fission time into the time the system needs to overcome the barrier (pre-saddle, $t_{presad}$) and the time it takes to descent to scission (post-saddle, $t_{postsad}$). This unfolding is important for getting insight into properties such as fission barriers, level densities, nuclear viscosity, etc. The critical role of entrance-channel asymmetry, on one side, and CN excitation energy and fissility, on the other side, in extracting the different times was demonstrated in Ref.~\cite{saxena:1993}. A recent study \cite{shareef:2016} pointed the still unresolved puzzle.\\
The un-ambigous extraction of fusion-fission time scales, and thus reliable conclusions on underlying nuclear properties, is a multi-dimensional problem. The fusion process depends on the projectile-target combination and the bombarding energy with respect to the Coulomb barrier ($E_{cm}$/$V_b$). These two features determine the angular momentum of the composite system. As for the fission stage, CN fissility $\chi$, excitation energy and angular momentum all play a role. A separate study of each effect is very challenging, if possible at all. Indeed, in a heavy-ion collision, these variables are usually not independent. To quote some example: Producing a compound nucleus with specific fissility and excitation energy by means of different entrance channels implies, in general, different bombarding energies, and leads to different angular momentum distributions for the CN. On the other hand, fixing $E_{cm}$/$V_b$ and $L$ usually requires varying $E^*$ and  $\alpha$/$\alpha_{BG}$. The involved multi-dimensional nature of the challenge is illustrated in fig.~\ref{fig1} where a representative set of experimental neutron pre-scission multiplicities $M_n^{pre}$ is shown, as a function of  $\alpha$/$\alpha_{BG}$ and $E^*$ in a), and as a function of $\chi$ and $E^*$ in b). We chose to consider neutron multiplicities here, as they were identified as relevant signatures of nuclear reaction times. Figure~\ref{fig1} demonstrates that, while $E^*$ seems to have the leading role in determining $M_n^{pre}$, entrance-channel asymmetry and CN fissility matter also.  We note that, on top of the displayed correlations, the angular momentum can vary as well. Furthermore, the comparison of the two panels makes it clear that an independent variation of each parameter is, in general, not available.\\

In their reference work, Saxena et al. \cite{saxena:1993} attempted to extract the individual contributions, $t_{fo}$, $t_{presad}$ and $t_{postsad}$, to the total fusion-fission time, by means of a systematics analysis of available experimental information on $M_n^{pre}$. Complete separation could not be achieved, but a definite correlation between specific time scales and, either $\chi$ or $E^*$, was observed. Shareef et al. \cite{shareef:2016} recently revisited the issue of ordering parameter, with focus on the significance of $\alpha$/$\alpha_{BG}$. From these and similar analyses \cite{hinde:1992}, it is clear that a unique ordering parameter is impossible to extract due to the multi-dimensional nature of the problem.\\
Since nuclear reaction times cannot be measured directly, model calculations are needed to link them to experimental observables, {\it i.e.} $M_n^{pre}$ in present case. In the aforementioned studies \cite{saxena:1993, hinde:1992, shareef:2016} the fusion stage is not explicitly modeled, and the fission stage is treated within the statistical model. Although the latter does not involve times as such, it is possible to implement the influence of time by some {\it ad hoc} ansatz (see Ref.~\cite{jurado:2007} for a recent discussion).\\
Describing the evolution of the system during the fusion stage, and its subsequent decay by fission, implies involved model calculations. Shape evolution as function of time is to be obtained from the solution of a multi-dimensional equation of motion, for both the (entrance) fusion and (exit) fission stages. Solving this equation is rather complex for each stage on its own already. Further coupling of the two stages is also particularly difficult, both in terms of optimal description of the shape and of the potential energy (we refer to Ref.~\cite{karpov:2017} for a update discussion on this challenge). In the present context and for the present goal, a fully-dynamical description does not exist yet. So far, hybrid models are used. Depending on the aim of the study, the dynamical treatment is restricted to, either the entrance \cite{wilczynska:1993, kumar:2014} or the exit \cite{frobrich:1998, adeev:2005}, stage. The present work is trying to go a step further in this direction: By means of dynamical considerations for both the entrance and exit channels, we attempt to deepen previous studies \cite{saxena:1993, shareef:2016} based on the statistical model. 

\section{Method and models}

\subsection{Strategy}

The characteristics of the reaction (projectile, target, bombarding energy) give the CN maximum excitation energy $E^*$ and angular momentum $L$. Though, even with initial $E^*$ and $L$ fixed, different entrance channels have shown to experience different decays \cite{ruckelshausen:1986}, what suggested different patterns for the entrance-channel relaxation in shape, excitation energy and angular momentum. Consequently, the CN formation (equivalently, fusion) time is different. What happens to the system during this time defines the initial conditions for the CN decay calculation, which it may affect more or less sizeably depending on the specificities of the entrance-channel.\\
To address the issue of influence of entrance-channel effects on typical fission observables, the following method is proposed. The fusion dynamics is calculated with the dynamical heavy-ion collision code HICOL. More specifically, the code is used to estimate the CN formation time as depending on the nuclear reaction at work. The fission stage is calculated from a multi-dimensional dynamical stochastic Langevin code. In this framework, pre- and post-saddle dynamics emerge naturally from the solution of the equation of motion, avoiding the ansatz required in statistical models. The dependence on the properties of the compound system (mass $A$, charge $Z$, $E^*$ and $L$), and their evolution with time, are consistently taken into account along the decay. The model is supposed to yield a fair description of the CN decay by fission, provided that a fully equilibrated and compact CN is formed in the entrance channel. Any discrepancy with experiment may thus be ascribed to an entrance-channel effect.\\
Combining the time information from HICOL with the CN-decay stochastic Langevin mode, we try to pin down the entangled influence of entrance-channel asymmetry, fusion time, CN fissility, excitation energy and angular momentum on the experimental observables.\\           
The goal of this study is to investigate, and possibly, unfold, entrance- and exit-channel aspects. Though, it is emphasized that the present work does {\it not} provide us with a complete, unified dynamical model of fusion-fission. Dynamical codes are run separately for the entrance (HICOL) and exit (stochastic Langevin) channels. Also, the outcome of the fusion code is not explicitly fed to the subsequent treatment of CN decay; it is used to estimate the magnitude of the fusion time, only. The investigation on the magnitude of entrance-channel effects remains thus qualitative at some level. However, the proposed method permits to re-examine experimental information in a multi-dimensional space, where previous work concentrated on one-dimensional (1D) projections for specific variables of the problem. The study shows that the influence of one or the other variable critically depends on the mass region under consideration. As a consequence, 1D projections can, either magnify or hide, the importance of some variable, depending on the system. That can lead to hazardeous interpretation when extrapolation is made in other mass regions where the dominating factor may have changed. In this context, the multi-dimensional analysis shall help to address the apparent unconsistency in measured neutron pre-scission multiplicities noted recently by Shareef et al. \cite{shareef:2016}.

\subsection{Theoretical framework}\label{model}

As introduced above, we do not embark on a unified model of fusion-fission, from the heavy-ion collision to the formation of cold fission fragments. Instead, we divide the description into two parts.  As for the first (fusion) stage, the HICOL \cite{feldmeier:1987} code is employed. The second (fission) stage is modelled with the stochastic Langevin code developed in Omsk \cite{adeev:2005, nadtochy:2014}. The predictions by HICOL are analyzed in terms of CN formation time, while we exploit the outcome of the fission stage in details. The HICOL information is used to interpret the difference, if any, between the predictions of the CN-decay calculation and experiment. The main ideas and ingredients of the aforementioned two theoretical codes are given below; we refer to the quoted references for further details.

\subsubsection{The HICOL code for fusion dynamics}

The HICOL code \cite{feldmeier:1987} gives access to the time evolution of shape relaxation in a heavy-ion collision by means of  a deterministic description of the motion. The collision is considered in the configuration space of three shape variables, representing, respectively, the distance between the geometrical centers of the reaction partners, the thickness of the neck connecting them, and the asymmetry in their size (or mass). Classical equations of motion of the Langevin type with inclusion of one-body dissipation in the form of the "wall and window" formula \cite{blocki:1978} are solved numerically in the 3D configuration space. The potential energy landscape is purely macroscopic, and given from the FRLDM liquid drop formula \cite{krappe:1979} accounting for the finite range of nuclear forces. Inertia is obtained in the Werner-Wheeler approximation of incompressible irrotational flow \cite{davies:1976}. HICOL does not contain free parameters and has shown to consistently describe the dynamical evolution of various composite systems formed in nucleus-nucleus collisions in a wide range of impact parameters. In the present work, we use the version of the code as implemented in Ref.~\cite{wilczynska:1993}. In the framework of the present work, it is important to note that particle evaporation during the path to fusion is {\it not} included in HICOL. 

\subsubsection{Stochastic Langevin approach to fission dynamics}

The model used to compute the heavy CN decay is based on the stochastic classical approach of nuclear dynamics \cite{abe:1996}. The four-dimensional (4D) Langevin code developed in Omsk, see Refs.~\cite{adeev:2005, nadtochy:2014}, is employed.\\
In the stochastic approach of fission, most relevant degrees of freedom are considered as collective coordinates, and their evolution with time is treated as the motion of Brownian particles, which interact stochastically with a surrounding "heat bath". In the present model, four collective coordinates are considered. Three variables $({\bf q})$ describe the shape of the deforming nucleus, and the fourth one corresponds to the orientation of its angular momentum relative to the symmetry axis. The shape coordinates are related to elongation, neck constriction, and left-right asymmetry, while the projection $K$ of the total angular momentum onto the symmetry axis of the fissioning nucleus is chosen for the fourth so-called tilting coordinate. The evolution of the shape and $K$ coordinates with time is obtained by solving, in parallel, the corresponding Langevin equations of motion, and assuming that the motion in the $K$ direction is over-damped. The driving potential is given by the Helmholtz free energy $F({\bf q},K)=V({\bf q},K) - a({\bf q}) T^2$, with $V({\bf q},K)$ being the potential energy, $a({\bf q})$ the level-density parameter \cite{ignatyuk:1975} and $T$ the temperature of the system. The potential energy $V({\bf q},K)$ is calculated within the framework of the macroscopic FRLDM model \cite{krappe:1979}. The calculation of inertia uses the Werner-Wheeler approximation \cite{davies:1976}, and friction is derived assuming the chaos-weighted one-body dissipation formalism \cite{chaudhuri:2001}, which can be considered as an improved variant of the "wall and window" prescription. De-excitation by evaporation of light particles  ($n$, $p$, and $\alpha$) by the compound system prior scission, as well as by the fragments after scission, is taken into account employing the Monte-Carlo approach. Particle-decay width are calculated within the Hauser-Feschbach theory. The 4D model of Omsk has shown able to explain a large variety of observables for fission over a wide range of systems (see Ref.~\cite{nadtochy:2014} and therein) \footnote{The unconsistency pointed in \cite{mahata:2017} was recently solved in the code, and found to have no impact for the concern of the present work.}.\\
We note that all Langevin calculations presented in this work were obtained using the prescription of Ref.~\cite{frobrich:1998} for modelling the CN initial $L$ distribution. The latter was compared to coupled-channel calculations by the CCFULL code \cite{hagino:1999, shrivastava:2017}. The maximum angular momenta could differ by several units, but the most probable $L$'s from Ref.~\cite{frobrich:1998} and CCFULL were found similar. Most important for the  concern of the present investigation is the trend as a function of entrance-channel asymmetry ($\alpha$/$\alpha_{BG}$), and its evolution with system size ($\sim \chi$) and bombarding energy ($E_{cm}$/$V_b$), which were observed to be the same for the two theoretical $L$ distributions.\\
\\
We note the consistency of the present framework. Both the models employed for treating the fusion and fission stage are based on macroscopic concepts, classical equations of motion, and similar prescriptions for model ingredients such as potential energy, friction, and inertia. We note also that all necessary information about the dynamical evolution till the CN is formed is available from HICOL, and can in principle be supplied as an input to the stochastic Langevin code. As explained above, this is not done here. As a first step, the strategy is to model the fission decay dynamics under Bohr's hypothesis, and to consider {\it a posteriori} the possible influence of fusion dynamics, guided by HICOL predictions. The main goal is to decouple entrance- and exit-channel dynamical effects.

\section{Results}

The reaction systems studied in this work are summarized in Table 1. Both lowly- and highly-fissile CN are investigated, located either below or above the BG point, from near to above Coulomb barrier energies. We restrict to reactions for which $E^*$ exceeds about 40 MeV, where the models used in this work are valid. The population in angular momentum ranges from about 10 to 80 $\hbar$. For those reactions for which experimental information on the neutron pre-scission multiplicity exists, the corresponding reference is specified in the table. We focus the investigation on $M_n^{pre}$ as it is particularly sensitive to fission time scales, and numerous data exist. Other observables are possible \cite{kumar:2014, kaur:2015}, and provide complementary insight. We note that most of the systems which we consider have been discussed also within a statistical model framework in {\it e.g.} Refs.~\cite{saxena:1993, shareef:2016}.

\begin{table}[h!]
\caption{Properties of the reactions studied in this work.}
\begin{tabular}{c|c|c|c|c|c}
\hline
Reaction &$E_{lab}$&$E^*$  &   $\alpha$/$\alpha_{BG}$   &  $\chi$   & Expt.\\
 &(MeV)&(MeV) &    &    & ref.\\
\hline
$^{28}$Si+$^{134}$Ba $\to$ $^{162}$Yb&132-165&81-115&0.85&0.614& - \\
\hline
$^{20}$Ne+$^{159}$Tb $\to$ $^{179}$Re&89-205&58-174&0.97&0.648& \cite{cabrera:2003}\\
\hline
$^{16}$O+$^{182}$W $\to$ $^{198}$Pb &72-80&45-53&1.00 & 0.7044& \cite{newton:1988}\\
$^{28}$Si+$^{170}$Er $\to$ $^{198}$Pb  &113-139&57-83&0.85 & 0.7044& \cite{mahata:2006}\\
\hline
$^{12}$C+$^{194}$Pt $\to$ $^{206}$Po &80-108&63-92&1.04& 0.716& \cite{golda:2013}\\
$^{22}$Ne+$^{184}$W $\to$ $^{206}$Po &99-127&63-92&0.93&0.716& - \\
\hline
$^{16}$O+$^{197}$Au $\to$ $^{213}$Fr &83-113&52-80 &0.99& 0.7428&  \cite{newton:1988}\\
$^{19}$F+$^{194}$Pt $\to$ $^{213}$Fr &82-111&50-80 &0.95 &0.7428& \cite{singh:2012}\\
\hline
$^{12}$C+$^{204}$Pb $\to$ $^{216}$Ra &86-90&57-61&1.02 & 0.7504 & \cite{singh:2008}\\
$^{19}$F+$^{197}$Au $\to$ $^{216}$Ra &93-97&57-61&0.95& 0.7504 & \cite{singh:2008}\\
\hline
$^{11}$B+$^{237}$Np $\to$ $^{248}$Cf &73-96&60-82&1.01&0.825& \cite{saxena:1993}\\
$^{16}$O+$^{232}$Th $\to$ $^{248}$Cf &96-118&60-82&0.97&0.825& \cite{saxena:1993}\\
 \hline
\hline
\end{tabular}
\end{table}

\subsection{Overview of the calculated results}\label{illus}

\begin{figure*}[!hbt]
\hspace{-1.4cm}
\includegraphics[width=20cm, height=7.cm, angle=0]{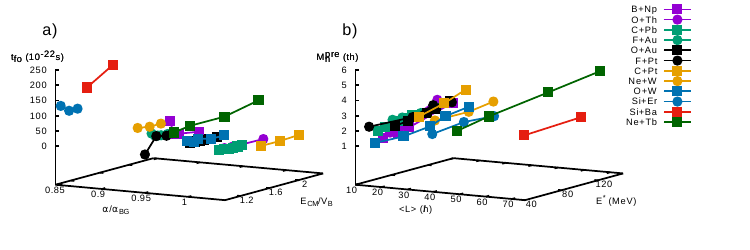}
\vspace{-.70cm}
\caption{(Color online) a) Calculated dependence of the formation time $t_{fo}$ on entrance-channel asymmetry $\alpha$/$\alpha_{BG}$ and energy with respect to the Coulomb barrier $E_{cm}$/$V_b$, as obtained from HICOL. b)  Calculated dependence of the neutron pre-scission multiplicity $M_n^{pre}$ on CN excitation energy $E^*$ and mean angular momentum $<L>$ as obtained from the stochastic Langevin code. The reactions presented in the plots are listed in Table 1.  Lines connect different energy points for a given reaction. The same color is employed for reaction couples leading to the same CN.} 
\label{fig2}
\end{figure*}

\begin{figure*}[!hbt]
\hspace{-0.85cm}
\includegraphics[width=16cm, height=6.4cm, angle=0]{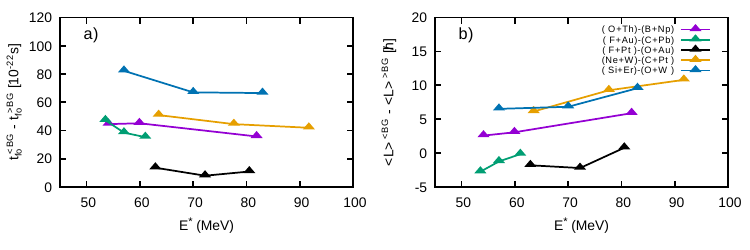}
\vspace{-.40cm}
\caption{(Color online) a) Difference between the formation times $t_{fo}$ (integrated over $L$) below ($< BG$) and above ($> BG$) the BG asymmetry $\alpha_{BG}$ as function of $E^*$ for the reaction couples leading to a given CN and considered in Fig.~\ref{fig2}. b) Similar to panel a) for the difference between the mean angular momenta $<L>$.} 
\label{fig2b}
\end{figure*}

A typical outcome of the calculations performed in this work is shown in Fig.~\ref{fig2}, as obtained, respectively, for the fusion andthe fission stage by the corresponding code. The correlation predicted by HICOL between CN formation time, entrance-channel asymmetry $\alpha$/$\alpha_{BG}$, and center-of-mass energy with respect to the Coulomb barrier $E_{cm}$/$V_b$ is shown in panel a). As introduced above, the projectile-target combination plays a key role in determining the time scale of the entrance-channel dynamics: for a given CN, the projectile-target couple with $\alpha < \alpha_{BG}$ needs more time to fuse than the combination with  $\alpha > \alpha_{BG}$. The violence of the collision, in terms of $E_{cm}$/$V_b$, plays some role as well. Panel b) displays the dependence of the neutron pre-scission multiplicity $M_n^{pre}$ on CN initial excitation energy $E^*$ and angular momentum $L$ as anticipated from the stochastic approach to fission. The picture precisely illustrates the features of the reaction mechanism which are accounted for in the dynamical CN-decay model used in this work. That is, the relative importance of $E^*$ and $L$, and its dependence on the CN ($A$, $Z$) composition. The calculated multiplicity is observed to be primarly governed by $E^*$. Though, for lowly-fissile systems, as we will see in more detail below, the angular momentum can play a non-negligible role.\\

A quantitative estimate of the aforementioned differences and related to the entrance channel is given in Fig.~\ref{fig2b}. For reaction couples yielding the same CN, the difference between the formation times $t_{fo}$ for the projectile-target combination located below and above $\alpha_{BG}$ is shown in panel a) as function of $E^*$. Similarly, panel b) displays the difference in mean angular momentum between reactions with $\alpha < \alpha_{BG}$ and $\alpha > \alpha_{BG}$. Based on the left panel, one may expect an enhanced  influence of the fusion stage for the couple $^{16}$O+$^{182}$W {\it vs.} $^{28}$Si+$^{170}$Er leading to $^{198}$Pb, as compared to the couple $^{12}$C+$^{204}$Pb {\it vs.} $^{19}$F+$^{197}$Au leading to $^{216}$Ra. The right panel suggests the possible emergence of $L$-driven entrance-channel dependences if the difference in $<L>$ for two reactions of a given couple is large enough. The differences constructed in Fig.~\ref{fig2b} are explored in detail further below.\\

The correlations displayed in Fig.~\ref{fig1} for experiment, and Fig.~\ref{fig2} for theory, and their concomitant analysis, are at the centre of this work. According to the right panel of  Fig.~\ref{fig2}, for the same $E^*$ and close-by $L$ values, two reactions forming the same compound are expected to yield similar pre-scission multiplicities, in full accordance with Bohr's hypothesis which the dynamical CN-decay calculation used here is based on. However, Fig.~\ref{fig1} shows that rather distinct  $M_n^{pre}$ were measured for several such couples of reactions. Inspection of the left panel of  Fig.~\ref{fig2} suggests that two reactions within a couple can be characterized by noticeably different fusion times depending on $E_{cm}$/$V_b$. This difference in $t_{fo}$, as function of $E^*$, is explicitly shown in Fig.~\ref{fig2}a). Remarking that, in experiment, neutrons emitted along the path to fusion can in general not be discriminated from those emitted along the fission stage, the excess of measured $M_n^{pre}$ may be attributed \cite{saxena:1993, shareef:2016, wilczynska:1993, hinde:1992} to a large CN formation time. Remind that HICOL does {\it not} account for evaporation along fusion. Hence, a deficit in the "CN-decay stochastic Langevin based" prediction for $M_n^{pre}$ as compared to the measuremnt may sign emission during the fusion mechanism. As we will demonstrate in this work, while such a direct connection sounds realistic in many cases, the situation can turn out to be more complex in others. The reason is the aforementioned entangled interplay of {\it e.g.} un-matched $E_{cm}$/$V_b$ and/or $L$ variables for a given combination of $\chi$ and $E^*$. The quantitative importance of the mismatch critically depends on the compound nucleus mass and charge, ranging from nearly no consequence to sizeable compensation effects. The latter may lead to apparent inconsistencies \cite{shareef:2016}.\\

\begin{figure}[!hbt]
\hspace{-0.45cm}
\includegraphics[width=9.cm, height=6.2cm,angle=0]{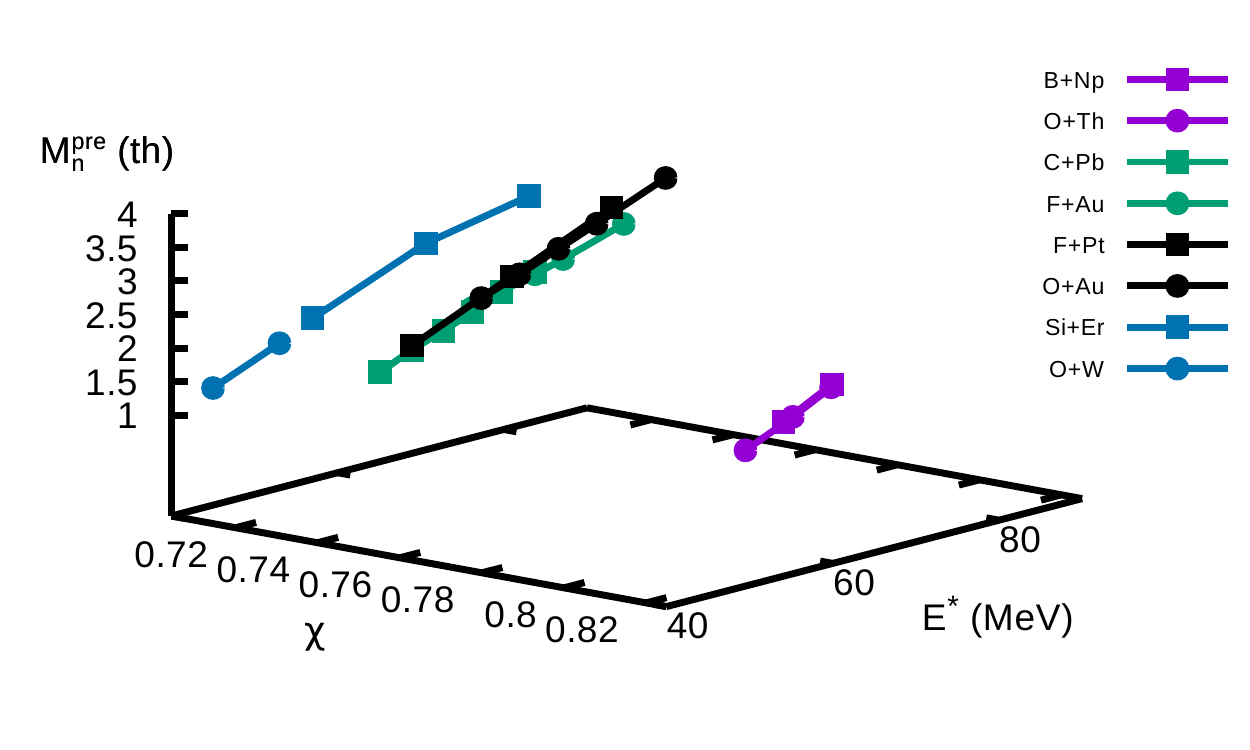}
\vspace{-.80cm}
\caption{(Color online) Calculated dependence of the neutron pre-scission multiplicity $M_n^{pre}$ on $\chi$ and $E^*$ as obtained from the stochastic Langevin code for the same set of reactions than in Fig.~\ref{fig1}. Lines connect different energy points for a given reaction. The same color is employed for reaction couples leading to the same CN.} 
\label{fig3}
\end{figure}

The previous paragraph establishes the strategy and ambition of the present work. A specific CN, characterized by {\it e.g.} $\chi$, can be populated at similar $E^*$ by means of different entrance channels. The latter govern the fusion time. For reactions with large values of $t_{fo}$ (basically, comparable to the decay time), light-particles can be emitted along the fusion process. Those are not calculated by HICOL, and are obviously not included in the Bohr's hypothesis framework used here for modelling CN decay. Figure~\ref{fig3} shows the same correlation as in Fig.~\ref{fig1}b), but as expected from the dynamical CN-decay code. Naturally, for a given CN, the theoretical results merge for different entrance channels, and exhibit a smooth and continuous evolution with $E^*$. Evidently, experiment, Fig.~\ref{fig1}b), contradicts the theoretical picture in most cases (deliberately shown). Guided by HICOL predictions, Fig.~\ref{fig2}a), we try to interpret the difference between measurements and Langevin CN-decay calculations.  Such an analysis is proposed to reveal, and hopefuly unfold, the possible influence of the first stage of the reaction on the interpretation of the second one. 

\subsection{Detailed analysis and comparison with experiment}\label{analysis}

\begin{figure*}[!hbt]
\includegraphics[width=16.cm, height=10.cm, angle=0]{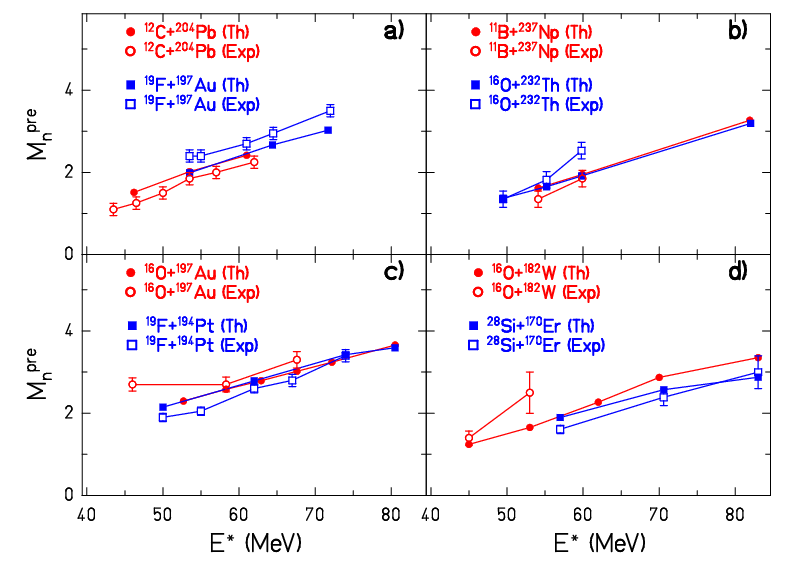}
\vspace{-.50cm}
\caption{(Color online) Neutron pre-scission multiplicity $M_n^{pre}$ as function of excitation energy for various reactions. Each panel corresponds to a given CN: a) $^{216}$Ra, b) $^{248}$Cf, c) $^{213}$Fr, and d) $^{198}$Pb, populated by means of two entrance channels. The experimental $M_n^{pre}$ (open symbols) are compared with dynamical Langevin CN-decay calculations (full symbols). For each couple of reactions, the system with largest (smallest) $\alpha$/$\alpha_{BG}$ is shown in red (blue).} 
\label{fig4}
\end{figure*}

Towards the aim of the work, we discuss in detail separately several representative reactions. The neutron pre-scission multiplicity expected from the dynamical Langevin CN-decay calculation for four compound nuclei is shown in  Fig.~\ref{fig4} as function of excitation energy. Each panel corresponds to a given CN: a) $^{216}$Ra, b) $^{248}$Cf, c) $^{213}$Fr, and d) $^{198}$Pb, populated by means of two entrance channels. Wherever available, experimental results (open symbols) are overlaid with theory (full symbols). As much as possible, we favored systems for which the two reactions were studied in the same paper, measured and analyzed under identical conditions, in order to minimize the possibility of experimental bias. {\it This is important for the present investigation which relies mostly on the relative difference in $M_n^{pre}$ between two reactions, rather than on absolute values.}\\

Starting with $^{216}$Ra, produced in either $^{12}$C+$^{204}$Pb or $^{19}$F+$^{197}$Au collisions, the calculation exhibits nearly the same $M_n^{pre}$ at overlaping $E^*$. This is due to the very similar $L$ distribution for the two reactions (see Fig.\ref{fig2b}b) where the difference in $<L>$ is very small). While theory explains reasonably well the reaction $^{12}$C+$^{204}$Pb within the experimental error bars, it predicts a slightly too small number of pre-scission neutrons for $^{19}$F+$^{197}$Au. Most noteworthy is that, independent on the description of absolute values, theory does not explain the difference seen in experiment between the two reactions. Having excluded strong $L$ effects, we thus ascribe this difference to the fusion stage. According to our calculations with HICOL, the CN formation time is, for the here-discussed energy range, larger by about a factor of 2 for $^{19}$F+$^{197}$Au ($\alpha$/$\alpha_{BG} < 1$) as compared to  $^{12}$C+$^{204}$Pb ($\alpha$/$\alpha_{BG} > 1$) at given $L$. At $E^* \approx$ 61 MeV the difference in $t_{fo}$ varies from 20 to 50 $^.$ 10$^{-22}$ s over the $L$ range populated, see Fig.\ref{fig2b}a). This is comparable to typical neutron emission times for heavy nuclei at corresponding excitation energy. Hence, we propose that the difference of about 0.5 neutron between the two systems is due to emission during the CN formation stage. Singh et al. \cite{singh:2008} suggested some influence of angular momentum, but also clearly mentionned entrance-channel dynamics.\\
Switching to $^{248}$Cf formed in $^{11}$B+$^{237}$Np and $^{16}$O+$^{232}$Th collisions \footnote{Note that in Ref.~\cite{shareef:2016} there is a mistake for this couple of reactions, $^{237}$Np having been replaced with $^{232}$Np. This is certainly due to the typo in the table of Ref.~\cite{saxena:1993} from where it was extracted. Hence, the point corresponding to B+Np in Fig. 4 of Ref.~\cite{shareef:2016} has $\alpha$/$\alpha_{BG}$=0.97 rather than 1.01.}, the calculation in Fig.~\ref{fig4} again suggests no dependence of $M_n^{pre}$ on the entrance channel at similar $E^*$. In that case, however, the calculated initial CN angular momentum distributions show quite some difference for the two projectile-target combinations. Similarly to Ref.~\cite{kumar:1994} we find that $^{16}$O+$^{232}$Th populates up to larger $L$'s (see Fig.\ref{fig2b}b) for $<L>$). The fact that the calculated $M_n^{pre}$ is not sensitive to the angular momentum is explained by the high fissility of the CN: the fate of $^{248}$Cf is not hindered by neutron emission, the fission barrier is much smaller than the neutron binding energy, and the system is commited to fission anyhow \cite{charity:1986}. As a consequence, for $^{248}$Cf like for $^{216}$Ra, the difference in the measured pre-scission multiplicities for the considered projectile-target combinations is classified as an entrance-channel effect. This is supported by the difference in $t_{fo}$ obtained from HICOL and in accordance with previous work \cite{kumar:1994}: the more symmetric reaction $^{16}$O+$^{232}$Th ($\alpha$/$\alpha_{BG} < 1$) has a non-negligible probability for evaporating a neutron during its longer fusion phase as compared to the more asymmetric  $^{11}$B+$^{237}$Np ($\alpha$/$\alpha_{BG} > 1$) combination.\\
The compound nucleus $^{213}$Fr considered in Fig.\ref{fig4}c) was synthetized in $^{16}$O+$^{197}$Au and $^{19}$F+$^{198}$Pt collisions. The calculated $M_n^{pre}$ shows negligible dependence on the reaction, as a result of the aforementionned two features, {\it i.e.} similar $L$ distributions, see Fig.\ref{fig2b}b), and a rather high fissility for the CN. The HICOL-predicted difference in formation time is rather small also, see Fig.\ref{fig2b}a), in contrast to the previous cases. In experiment, there is finite difference in  $M_n^{pre}$. Most notably is the fact that $^{16}$O+$^{197}$Au ($\alpha$/$\alpha_{BG} \approx 1$) displays a larger $M_n^{pre}$ as compared to $^{19}$F+$^{198}$Pt ($\alpha$/$\alpha_{BG} < 1$). That is, the difference in $\alpha$/$\alpha_{BG}$ (equivalently, in $t_{fo}$) does {\it not} reflect the measured $M_n^{pre}$ values; it is even opposite to expectation \cite{shareef:2016}. In an attempt to understand this observation, we first note that the difference in $t_{fo}$ for the close F+Pt and O+Au couples is much less than for the combinations used in the $^{216}$Ra and $^{248}$Cf cases. Second, the difference in experimental $M_n^{pre}$ is seen to be limited. Third, we remark that the energies involved are sub- and near-barrier, what may imply additional effects \cite{mukarami:1986}. Finally, the two reactions were measured in different experiments and involved rather thick targets. Depending on the treatment of energy loss corrections, and the method for determining the suited excitation energy, the $x$-axis in the two experiments are subject to be shifted one with respect to the other by up to a few MeV. All in all, we think that one shall maybe not over-interpret the confrontation of the  F+Pt and O+Au data. The apparent, inconsistent \cite{shareef:2016}, difference in experimental $M_n^{pre}$ between the two entrance channels may be the outcome of a complex interplay between various weak physical effects, on on side, as well as, possible experimental bias and limited accuracy, on the other side.\\
The fourth compound considered in Fig.~\ref{fig4} is the less fissile $^{198}$Pb nucleus which pre-scission neutron multiplicities were measured in $^{16}$O+$^{182}$W and $^{28}$Si+$^{170}$Er fusion-fission reactions \footnote{Note that as compared to Ref.~\cite{shareef:2016} the $E^*$ values which we calculated for $^{28}$Si+$^{170}$Er are slightly higher, in accordance with Newton et al. \cite{newton:1988}, what yields a slightly shifted $E^*$ range.}. Only few, rather scattered, experimental points exist; some are around the barrier, have limited accuracy, and come from different experiments. Though, it is interesting to study this case as it reveals interesting features which the above fissile systems do not permit to study. The calculations were performed over an extended $E^*$ range for $^{16}$O+$^{182}$W in order to allow a more robust comparison with $^{28}$Si+$^{170}$Er. According to the scatter of some experimental points, and within the established precision of the model (see Refs.~\cite{adeev:2005, nadtochy:2014} and therein), the Langevin CN-decay calculation describes reasonably the measurements. A considerably larger formation time is predicted by HICOL for $^{28}$Si+$^{170}$Er ($\alpha$/$\alpha_{BG} < 1$), see Fig.~\ref{fig2b}a). One may thus anticipate a larger pre-scission multiplicity for this reaction. However, the measured $M_n^{pre}$ seems to be larger for $^{16}$O+$^{182}$W ($\alpha$/$\alpha_{BG} \approx 1$) in the overlaping $E^*$ region covered. This observation was left partly un-understood \cite{shareef:2016}. Beside the word of caution we emitted previously, it seems unlikely that experimental bias can alone explain why experimental observation is so far from what expected from the calculated sizeable large $t_{fo}$ value for the more symmetric reaction. We suggest that one reason why the multiplicities of the two reactions come close, and possibly are inversed as compared to expectation based on $t_{fo}$, is the significant influence of angular momentum. This is shown in Fig.~\ref{fig4}d) with the CN-decay calculations studied over an extended $E^*$ region. At given, low excitation energy, the calculated $M_n^{pre}$ is similar for $^{16}$O+$^{182}$W and $^{28}$Si+$^{170}$Er as seen from the close lying full circles and squares. With increasing energy the two curves separate due to an increasing difference in $<L>$ population between the two reactions (see Fig.~\ref{fig2b}b) for $<L>$). Below the bombarding energy leading to $E^* \approx$ 65 MeV, the most probable $L$'s contributing to fission differ by about 10 $\hbar$, while for the points above $E^* \approx$ 65 MeV, the $^{28}$Si+$^{170}$Er reaction populates angular momenta which on average exceed those from $^{16}$O+$^{182}$W by 15 $\hbar$ and more. For a CN of moderate fissility such as $^{198}$Pb, the barrier to overcome and proceed to fission is high at small $L$ (around 12 MeV) and becomes comparable to the neutron binding energy (around 9 MeV) at the highest $L \approx$ 55 $\hbar$ populated here. Hence, the competition between neutron emission and fission is particularly effective at large $L$, and any additional evaporated neutron can definitively prevent the system to go to fission. At the same time, the higher the angular momentum, the larger the excitation energy stored in rotational energy, the lower the intrinsic energy available for either evaporating neutrons or overcoming the fission barrier. For the points leading to $E^* \approx$ 65 MeV, the difference in maximal angular momentum populated in $^{16}$O+$^{182}$W and $^{28}$Si+$^{170}$Er collisions leads to 4 MeV less excitation energy for the latter system. This deficit in available intrinsic energy increases with increasing bombarding energ (equivalently, $L$), approaching the neutron separation energy. The $^{198}$Pb nuclei produced in Si+Er collisions have thus less intrinsic energy than when formed the O+W combination at the same total CN excitation energy $E^*$. Those $^{198}$Pb systems  formed in the Si+Er bombardment which finally succeed in overcoming the barrier are therefore those which emit less neutrons. That is at the origin of the progressive deviation of the two calculated curves in  Fig.~\ref{fig4}d). Angular momentum effects thus compensate formation time effects in determining the neutron pre-scission multiplicity. This may explain the trend which, despite uncertain at some level, seems to emerge from the experimental points. Nevertheless, it is not excluded that contribution from quasi-fission \cite{berriman:2001}, difficult to discriminate from fusion-fission experimentally, and characterized by lower reaction times, can partly explain the lower multiplicity for $^{28}$Si+$^{170}$Er.

\begin{figure}[!hbt]
\hspace{-1.1cm}
\includegraphics[width=9.6cm, height=12.3cm, angle=0]{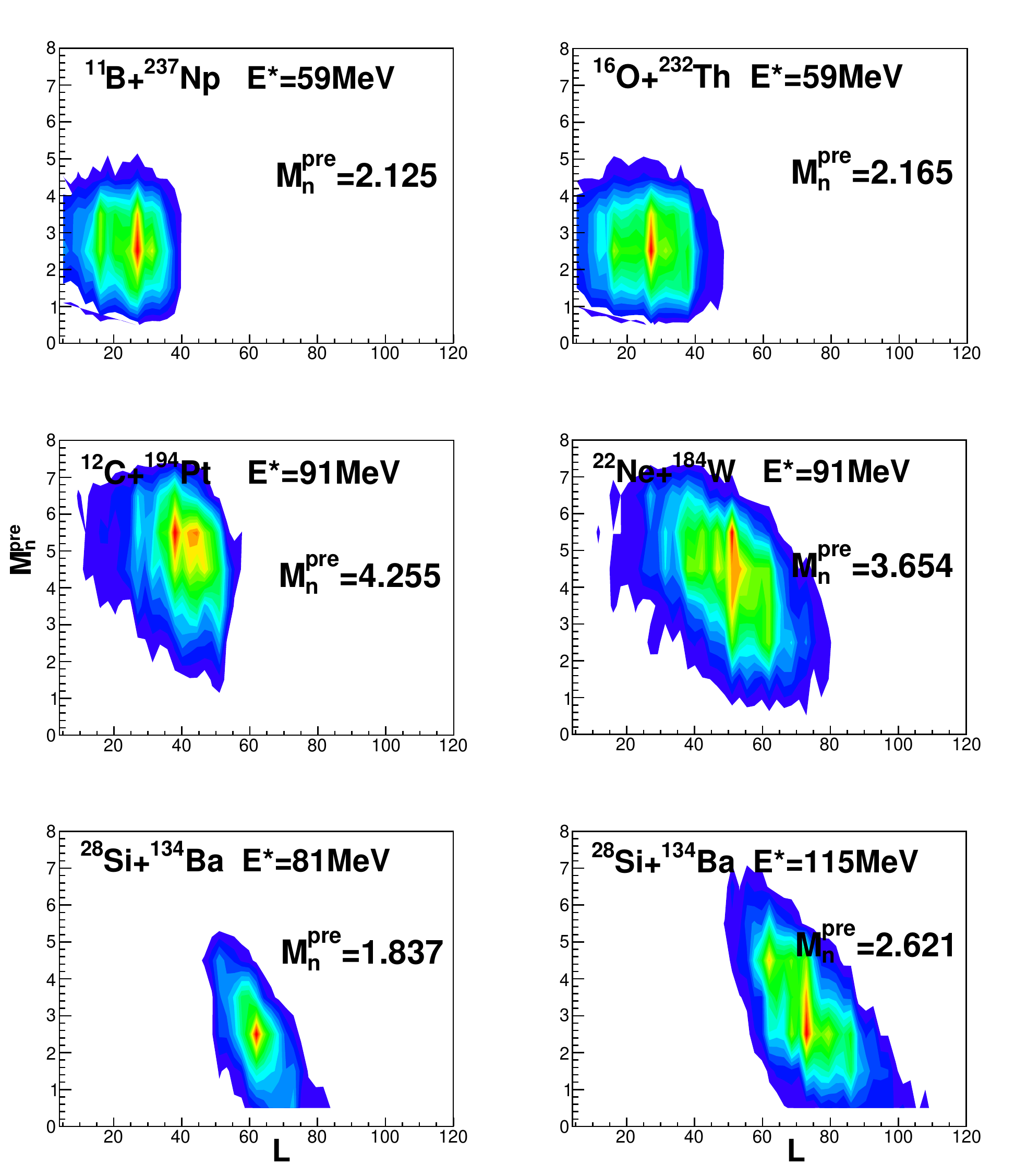}
\vspace{-.30cm}
\caption{(Color online) Correlation between the CN angular momentum $L$ and the neutron pre-scission multiplicity $M_n^{pre}$ as calculated from the stochastic Langevin CN-decay model. The first and second rows display each two reactions leading to a given CN ($^{248}$Cf and $^{206}$Po, respectively) at similar $E^*$. The last row considers a single reaction, at two different $E^*$. In each casel, the predicted mean $M_n^{pre}$ is indicated also.} 
\label{fig5}
\end{figure}

The complex picture that progressively develops with decreasing fissility in the CN-decay is investigated in Fig.~\ref{fig5}. The correlation between neutron pre-scission multiplicity and CN angular momentum calculated from the stochastic Langevin model is shown for systems with decreasing fissility ($^{248}$Cf, $^{198}$Pb, $^{162}$Yb, from top to bottom). The first and second rows each display a combination of reactions producing the same CN at similar excitation energy. The last row considers a single projectile-target combination at two bombarding energies. The calculated mean pre-scission multiplicity (after folding with the $L$ distribution) are indicated also. Confrontation with experiment is not possible for the second and third rows due to missing measurements. Though these systems are considered here as per deepening our investigation of entrance-channel dependence.\\
As discussed earlier, the neutron pre-scission multiplicity predicted by the CN-decay model is seen to be not sensitive to angular momentum for the fissile $^{248}$Cf compound; this is materialized in panels a) and b) of Fig.~\ref{fig5} by the slope of the mean $M_n^{pre}$ as function of $L$ which is nearly horizontal.\\
The observation made by comparing the $^{12}$C+$^{204}$Pb and $^{28}$Ne+$^{184}$W reactions leading to $^{206}$Po resembles the result detailed for $^{198}$Pb above: $M_n^{pre}$ on average decreases with increasing $L$. At given angular momentum, $M_n^{pre}$ is the same for both entrance channels. The difference in the calculated mean multiplicities, indicated in each panel, is then due to the different CN angular momentum population in the two reactions, as obvious from Fig.\ref{fig5}, combined to the crucial influence of $L$ in this mass region. For that particular couple, it leads to a lower pre-scission multiplicity  for the fission stage for $^{28}$Ne+$^{184}$W. According to HICOL, the $^{28}$Ne+$^{184}$W ($\alpha$/$\alpha_{BG} < 1$) reaction needs, however, more time than $^{12}$C+$^{204}$Pb ($\alpha$/$\alpha_{BG} > 1$) to produce a fully equilibrated the CN, and neutrons may be emitted during this phase. That would increase the measured pre-scission multiplicity for $^{28}$Ne+$^{184}$W as compared to the prediction from the CN-decay calculation. Depending on the ratio of neutrons emitted, respectively, during the CN formation and CN decay stage, fusion time and angular momentum effects can act in the same or opposite direction in influencing the pre-scission multiplicity for $^{28}$Ne+$^{184}$W. The latter would then be smaller or higher than for $^{12}$C+$^{204}$Pb, and it may be classified as, either consistent or unconsistent \cite{shareef:2016}. The present schematic method does not permit to determine quantitatively which, from the two stages, will dominantly contribute in this particular case. However, it illustrates again that the direction of various effects caused by $\alpha$/$\alpha_{BG}$, $L$, $E_{cm}$/$V_b$, and the magnitude of their interplay, can be complex to un-fold from the sole measurement of neutron pre-scission multiplicities. Nuclei at and below $A \approx$ 180-200 show particularly involved features \cite{gontchar:2004, ryabov:2008, charity:1986}. As in previous studies \cite{gontchar:2004}, we observe large fission times in this region, what can further diminish the relative importance of the fusion time. Most measured neutrons are then accounted for by the CN-decay stage.\\
The last row of Fig.~\ref{fig5} considers the reaction $^{28}$Si+$^{134}$Ba leading to $^{162}$Yb at excitation energies of $E^ * =$ 81 and 115 MeV. In this light system, the fission barrier is larger than the neutron binding energy up to high $L$, and still additional features are anticipated \cite{gontchar:2004, ryabov:2008}. Very different angular momenta are populated at the two bombarding energies, and  $M_n^{pre}$ decreases very fast with $L$. In spite of the dominant contribution from the highest partial waves in fission of lowly-fissile systems and which would reduce the number of emitted neutrons similarly to the $^{206}$Po case discussed above, it is $E^*$ which finally determines the multiplicity for this even lighter compound nucleus. The reason is as follows. Due to the large fission barrier, the system has to overcome the saddle rather early along the cascade, when still a sizeable amount of $E^*$ is available. The fission time is indeed found to be reduced as compared to the  $A \approx$ 180-200 region, in accordance with the results of Ref.~\cite{gontchar:2004}. It may thus be conjectured that, for such light systems, the relative contribution from the fusion stage may again pick up, and again be more straightforwardly reflected in the measured multiplicity.

\begin{figure}[!hbt]
\hspace{-0.8cm}
\includegraphics[width=9.cm, height=5.8cm,angle=0]{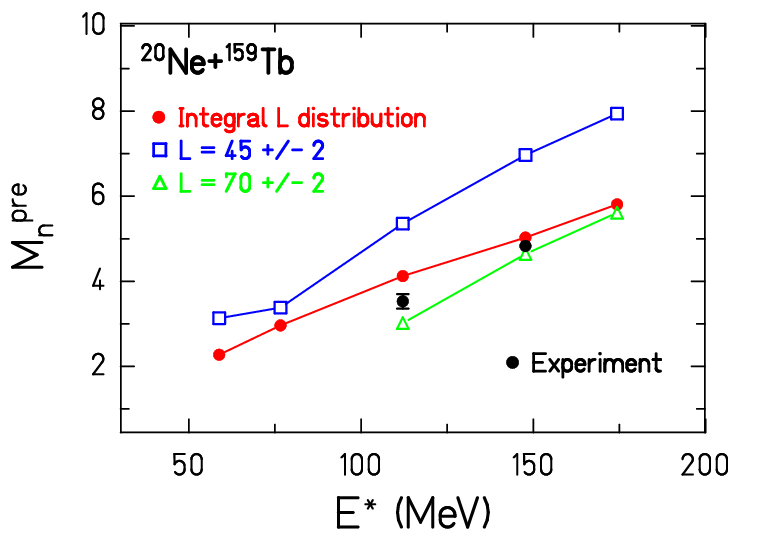}
\vspace{-.50cm}
\caption{(Color online) Neutron pre-scission multiplicity $M_n^{pre}$ as calculated from the stochastic CN-decay model for the  $^{20}$Ne+$^{159}$Tb reaction as function of CN excitation energy. The calculation, including the full theoretical $L$ distribution is shown with (red) dots, while calculations from selected $L$ values - 45 and 70 $\hbar$, are displayed with open (blue) squares and (green) triangles, respectively. Experimental data wherever available are shown as well \cite{cabrera:2003}.} 
\label{fig6}
\end{figure}

Finally, the effect of angular momentum in lowly-fissile systems is illustrated in Fig.~\ref{fig6} where the calculated pre-scission multiplicity for $^{20}$Ne+$^{159}$Tb leading to $^{179}$Re is shown as function of CN excitation energy. This system is intermediate between the above-discussed Ba and Po cases. In addition to the calculation folded with the entire $L$ distribution (which describes experiment where available \cite{cabrera:2003}), the result obtained in two narrow $L$ windows (45 and 70 $\hbar$) are displayed. Over the considered $E^*$ interval, the maximum CN angular momentum ranges from 60 to 100 $\hbar$. It is observed that, depending on the total $E^*$ excitation energy, the predicted multiplicity is given by a specific $L$ window, and which does not necessarily coincide with the highest populated CN partial waves. This is related to the complex competition between the influence of $E^*$ and $L$ and which is very sensitive to the CN fissility. That can blur the influence of the fusion stage. A proper unfolding of the different contributions is most challenging in the Pb region.

\section{Discussion}\label{discussion}

In this work, we have proposed a schematic method to address the possible critical influence of the fusion stage on the extraction of information about fission dynamics from heavy-ion collision experiments. That implies dealing with the influence of entrance-channel mass asymmetry, bombarding energy, angular momentum and excitation energy imparted to the compound nucleus. The interplay between all these paramaters can be complex, and possibily give rise to apparent inconsistencies \cite{shareef:2016} when a measurement is analyzed as function of one or the other parameter. The reason of this lies probably in the impossibility to determine a unique ordering paramater. In an attempt to"deconstruct" the multi-dimensional nature of the problem, we use a stochastic dynamical fission model, and compare its predictions in terms of neutron pre-scission multiplicity for various reactions. The relative difference in $M_n^{pre}$ between bombardments leading to the same CN, for such a calculation {\it vs.} experimental observation, is then discussed with predictions by the HICOL code for the fusion stage.\\
As emphasized above, in our exploratory investigation, among the various quantities computed by HICOL, we so far analyze only the fusion time, to explain possible discrepancy between experiment and dynamical CN-decay calculations. However, concomitant with the CN formation time, depending on the reaction,  HICOL also expects some saturation of the $E^*$ imparted to the CN, which is below the value given by the incident energy and fusion $Q$ value, as well as a hindrance to fusion for the highest $L$'s. The information from HICOL can be fully \cite{wilczynska:1993} or in part \cite{kumar:2014} used to compute the subsequent decay. This is not done here. Exploiting the richness of HICOL as an input to the CN de-excitation is beyond the scope of this work, and can be considered in future. We made few tests and obtained that such an improvement of the method can affect the quantitative result for some reactions. Nevertheless, it won't affect the qualitative output of this work, and the take-home message would remain the same. The main results obtained within the proposed framework are summarized below.\\  
Over the range of systems studied, measurements (Fig.~\ref{fig1}) like theory (Fig.~\ref{fig2}) show that $E^*$ plays a leading role in determining the neutron multiplicity. While experimental observations suggest that entrance-channel asymmetry plays the next key role in enhancing $M_n^{pre}$ when projecting data in a specific sub-space, the present work shows that this proposal is not the full story. Projection into other sub-spaces demonstrate namely the possible influence of angular momentum. The importance of either formation time or angular momentum at given $E^*$ critically depends on the fissility of the CN. Suming up, on top of the usually dominating influence of $E^*$, the measured $M_n^{pre}$ value is determned by a delicate balance governed by $\alpha$/$\alpha_{BG}$, $\chi$, and $L$. Depending on the magnitude of $E^*$, and on the $\chi$ region, the competition between $\alpha$/$\alpha_{BG}$ and $L$ is in favor of one of the other parameter. While the former usually dominates in fissile systems, the latter can be hidden behind not-necessarily intuitive observations for CN with masses around and below $\approx$ 200. Thanks to the employed models, the present work allows projecting the puzzle onto various parameter sub-spaces, and this way, highlighting compensation effects. The latter may explain observations classified as un-resolved so far.\\
As noted in introduction, the present study does not pretend to provide a unified model for fusion-fission dynamics. In this sense, the schematic method does not guarantee robust quantitative results. However, it provides a suited framework for a reliable qualitative study. The multi-dimensional analysis permitted us to reveal the influence of the different ordering parameters which were proposed in the past, and which all had to face limitations for specific data. The complexity of the interplay between effects un-folded to some level in this work is proposed as a major reason for apparent inconsistencies.\\
We mention that the puzzle deconstructed in this work may be visible only for thoses events which end with fission, since the population in $L$ affects the fission probability. Observables connected to other competing channels (see Refs.~\cite{kumar:2014, ruckelshausen:1986} and therein) are therefore very useful also to complete the understanding.\\   
Finally, the present exploratory investigation suggests that, to  un-ambigously unfold the respective role of fusion and fission dynamics, and further un-bias the extraction of nuclear properties, requires still a large set of experimental data, spanning an as wide as possible mass domain.

\section{Conclusions}\label{summary}

In order to address puzzling experimental observations on fission observables from heavy-ion collisions, we combine a dynamical stochastic Langevin model for fission of excited compound nuclei, with predictions by the dynamical deterministic HICOL code for the fusion time. A careful study about the influence of entrance-channel asymmetry, bombarding energy, compound-nucleus fissility, angular momentum and excitation energy, is performed, and confronted to experimental results whenever available. Additional theoretical calculations spanning a wider domain in compound-nucleus are done, in order to reveal the complex picture that develops due to the possibility of compensation effects depending on the size of the system.  The work deepens previous analysis based on statistical models by the use of a description of the CN-decay which accounts for the dynamical features of the process and preserve the correlation between all quantities along time evolution in a natural way. Although model-dependent, and qualitative to some extent, the analysis performed in this work suggests that so-far unresolved observations can be understood within the multi-dimensional character of the process.\\
\\
\\
{\bf Acknowledgements}
\\
We thank Pr.~K. Siwek-Wilczynska and Dr. A.~Shrivastava for help in HICOL and CCFULL calculations, respectively. We are very grateful to Dr. A.~Chatterjee for discussion related to experimental and statistical-model analysis, and to Pr. Adeev for careful reading of the manuscript. The work was sponsored by the French-Polish agreements LEA COPIGAL (Project No.~18) and IN2P3-COPIN (Project No.~12-145).

\end{document}